\documentclass[aps,twocolumn,showpacs]{revtex4}
\usepackage{graphicx}
\usepackage[all]{xy}
\usepackage{amsmath}
\usepackage{amssymb}
\newcommand{\be}{\begin{equation}}
\newcommand{\ee}{\end{equation}}
\newcommand{\ben}{\begin{eqnarray}}
\newcommand{\een}{\end{eqnarray}}
\newcommand{\bes}{\begin{subequations}}
\newcommand{\ees}{\end{subequations}}

\newcommand{\bb}{\bibitem}
\begin{document}
\title{Deformed Defects for Scalar Fields with Polynomial Interactions}
\author{D. Bazeia,$^a$ M.A. Gonz\'alez Le\'on,$^b$ L. Losano,$^a$ and J. Mateos Guilarte$^c$}
\affiliation{{\small $^a$Departamento de F\'{\i}sica, Universidade Federal da Para\'{\i}ba\\
Caixa Postal 5008, 58051-970 Jo\~ao Pessoa, Para\'{\i}ba, Brazil\\
$^b$Departamento de Matem\'atica Aplicada, Universidad de Salamanca, Spain\\
$^c$Departamento de F\'{\i}sica Fundamental and IUFFyM, Universidad de Salamanca, Spain}}

\begin{abstract}
In this paper we use the deformation procedure introduced in former work on deformed defects to investigate several new models for real scalar field.
We introduce an interesting deformation function, from which we obtain two distinct families of models, labeled by the parameters that identify the deformation function. We investigate these models, which identify a broad class of polynomial interactions. We find exact solutions describing global defects, and we study the corresponding stability very carefully.
\end{abstract}


\maketitle

\section{Introduction}
The study of defect structures is of great interest to high energy physics, and have been the subject of many interesting investigations, as we can see for instance in the books quoted in Ref.~{\cite{F}}. As one knows, defect structures are also of interest to condensed matter physics, and there they have been used to model different situations, as they appear for instance in Ref.~{\cite{M}}.   

In general, defect structures can be topological or non-topological, and in models described by a single real scalar field in $(1,1)$ space-time dimensions they are usually named kinks or lumps, respectively. They are global defects, and in this work we investigate kinks and lumps in models described by real scalar field, taking advantage of former work on deformed defects \cite{dd} to build new families of models, together with their corresponding defect solutions. Our investigation deals with the type-1 family of deformations, and we focus mainly on the basic properties of the models and defect structures, including stability, which is of direct interest to non-linear science in general.
The new models and solutions which we introduce in this work bring new interactions into the stage, and we believe that they are of interest both to condensed matter and high energy physics.

In condensed matter, scalar fields are important to describe spontaneous breaking of discrete symmetry, which may be a mechanism to describe the mass gap for fermionic carriers with Yukawa coupling to the bosonic degrees of freedom \cite{cm}. In high energy physics, scalar fields play central role in the standard model, controlling the way the masses of the elementary particles are generated under the Higgs mechanism \cite{SM}. Scalar fields also appear as bosonic portions of chiral superfields, leading to appropriate tool to implement supersymmetry in high energy physics \cite{susy}. They also play important role in string theory, in particular in type IIB superstring theory where the dilaton appears as a Ramond-Ramond scalar field which may control the string coupling; also, in string theory we should in general mention the many associated moduli fields, which are other examples of scalar fields playing important role \cite{S}. Furthermore, motivated by string theory, scalar fields may also be used to generate defect structures in noncomutative space-time \cite{snc} and in models which break Lorentz and CPT symmetries \cite{CKL,lb}. They can also play important role in supergravity \cite{CS}, and in braneworld \cite{b}, braneworld cosmology \cite{bc} and cosmology \cite{c}. In braneworld and in braneworld cosmology, scalar fields may constitute the driving mechanism to establish the problem, and in cosmology they may be used to source the dark energy needed to accelerate the Universe \cite{a}. For recent investigations of scalar field models in cosmology and in braneworld see, e.g., Refs.{\cite{cos,bra}}.

There are other instances where scalar fields play specific role. Some of them are related to the Peccei-Quinn (PQ) mechanism \cite{PQ}, which deals with the strong CP problem, and the Affleck-Dine (AD) mechanism \cite{AD}, which deals with baryogenesis -- for a review on the matter-antimatter asymmetry see Ref.~{\cite{ADR}}. Here we recall very recent investigations, in which if the scalar field potential is supposed to admit a tracker solution, the scalar field present in the PQ mechanism can also account for dark energy \cite{prl04}, and also in a variant of the AD mechanism, if one properly couples scalar field to neutrinos, one can induce net baryon number density in the model \cite{prl05}. Scalar fields may also provide important environments for semiclassical investigations in quantum field theory \cite{2dqft}. In spite of all the above possibilities, however,
neither condensed matter nor high energy physics suggests how the scalar fields interact or self-interact, although sometimes we know that they must self-interact to make up the model.

Scalar field may engender polynomial or non-polynomial self-interactions, and in the present work we focus on polynomial potentials. We use the deformation procedure of Ref.~{\cite{dd}} to introduce new families of models described by a single real scalar field, engendering different power of self-interactions, which we solve to find defect structures with the methodology for deformed defects set forward in these works. Other interesting investigations on polynomial potentials have been considered for instance in \cite{poly}, but there one hardly finds exact solutions for polynomial
potential of degree higher than six. In the present work we focus mainly on the basic issues, concerning the models and their exact solutions, and the corresponding stability. 

The main results of this investigation are of immediate interest to the braneworld scenario set forward in the last work in Ref.~{\cite{dd}}. The specific results concerning the presence of exact solutions impact directly on the former understanding, that it is hard to find defect solutions in models described by scalar field with polynomial interactions of higher and higher power in the field. For this reason, we believe that the present work is also of interest to the semiclassical program which computes corrections to kinks in quantum field theory in $(1,1)$ dimensions, in the case of cylinders and strips \cite{2dqft}. In the case of infinity length, the one-loop correction to the classical energy of these new deformed kinks might be computed using the generalized zeta function procedure proposed in Ref.~{\cite{iflg}}.

\section{Generalities}
 
We use the deformation procedure introduced in Ref.~{\cite{dd}} for real scalar fields in the standard, infinite volume case. We consider the two models
\bes
\ben
{\cal L}=\frac12\partial_\mu\chi\partial^\mu\chi-V(\chi)\label{1a}
\\
{\cal L}_d=\frac12\partial_\mu\phi\partial^\mu\phi-V(\phi)\label{1b}
\een\ees
where $\chi$ and $\phi$ are two real scalar fields and $V(\chi)$ and $V(\phi)$ are given potentials, which specify each one of the two models. We introduce a function $f=f(\phi),$ named deformation function, from which we link the model of (\ref{1a}) with the deformed model of (\ref{1b}) by relating the two potentials $V(\chi)$ and $V(\phi)$ in the very specific form
\be
V(\phi)=\frac{V(\chi\to f(\phi))}{(df/d\phi)^2}
\ee
This allows showing that if the starting model has finite energy static solution $\chi(x)$ which obeys the first-order equations
\be
\frac{d\chi}{dx}=\pm\sqrt{2V(\chi)}
\ee
and the equation of motion
\be
\frac{d^2\chi}{dx^2}=\frac{dV}{d\chi}
\ee
then the deformed model has finite energy static solution given by $\phi(x)=f^{-1}(\chi(x)),$ which obeys
\be
\frac{d\phi}{dx}=\pm\sqrt{2V(\phi)}
\ee
and
\be
\frac{d^2\phi}{dx^2}=\frac{dV}{d\phi}
\ee
The proof was already given in Ref.~{\cite{dd}}. We notice that if the deformation function $f(\phi)$ has critical points, the deformation procedure works smoothly if and only if the potential $V(f(\phi))$ has  zeroes of multiplicity at least two at those critical points. The critical points of $f(\phi)$ are the branching points of $f^{-1}(\chi),$ which is consequently a multivalued function. 

\section{New family of models}

We now proceed to the key point in the present work. To do this, we start with the standard ${\chi^4}$ model, and we search for new analytically solvable models, which support topological (kink-like) and/or non-topological (lump-like) defect structures as classical static solutions of the corresponding equations of motion. The methodology follows Ref.~{\cite{dd}} and especially the case of the type-1 family of deformations, that is, we search for new functions $f=f(\chi)$ from which we can build new families of models. The starting model is described by the potential
\be
U(\chi)=\frac12\;(1-\chi^2)^2
\ee
where we are using dimensionless field and coordinates. We fix the center of the defect at the origin ($x_0=0$) to get the kink-like solution
\be
\chi(x)=\pm\tanh(x)
\ee
We now choose a new deformation function
\be\label{df}
f(\phi)=\cos(a\arccos({\phi})-m\pi)
\ee
This function depends on $a,$ which is real constant, and on $m,$ which can be integer or semi-integer. The procedure follows \cite{dd} and
the parameter $m$ leads to two distinct families of models: for $m$ integer, the deformed potential can be written in the form
\be\label{vs1}
{V}_{\sin}^a(\phi)=\frac{1}{2a^2}(1-\phi^2)\sin^2\left(a\;\arccos\phi\right)
\ee
However, for $m$ semi-integer we get
\be\label{vc1}
{V}_{\cos}^a(\phi)=\frac{1}{2a^2}(1-\phi^2)\cos^2\left(a\;\arccos\phi\right)
\ee
As we show below, using half-integer and integer values for the parameter $a$ lead the number of vacua of the new model to be fixed at will.

The above models identify families of potentials which present static solutions given by
\be\label{sol1}
{\phi}(x)=\cos\left(\frac{\theta(x)+ m\;\pi}{a}\right)
\ee
where $\theta(x)$ is the principal determination of $\arccos(\tanh(x))$, e.g., $\theta\in [0,\pi]$.

An interesting aspect of these potentials is that they can be written in polynomial form, for $a$ integer or semi-integer, as follows
\be
{V}_{\sin}^a(\phi)=\frac{1}{4a^2}(1-\phi^2)(1-P_a (\phi))
\ee
and
\be
{V}_{\cos}^a(\phi)=\frac{1}{4a^2}(1-\phi^2)(1+P_a (\phi))
\ee
where
\be
P_a(\phi)=\sum_{k=0,2,..}^{2a}(-1)^{k/2}{{2a}\choose{k}}\phi^{2a-k}\left(1-\phi^2\right)^{k/2}
\ee
They are very interesting polynomial potentials, for which we can find all the static solutions explicitly.

We illustrate some of the many possibilities with the cases $a=1/2,3/2,5/2.$ Here we have, for the $sine$ family
\be
{V}_{\sin}^{1/2}(\phi)=\left(1-\phi^2\right)\left(1-\phi\right)
\ee
\be
{V}_{\sin}^{3/2}(\phi)=\frac{1}{9} \left(1-\phi^2\right)\left(1+3\phi-4\phi^3\right)
\ee
\be
{V}_{\sin}^{5/2}(\phi)=\frac{1}{25} \left(1-\phi^2\right)\left(1-5\phi+20\phi^3-16\phi^5\right)
\ee
and for the $cosine$ family
\be
{V}_{\cos}^{1/2}(\phi)=\left(1-\phi^2\right)\left(1+\phi\right)
\ee
\be
{V}_{\cos}^{3/2}(\phi)=\frac{1}{9} \left(1-\phi^2\right)\left(1-3\phi+4\phi^3\right)
\ee
\be
{V}_{\cos}^{5/2}(\phi)=\frac{1}{25} \left(1-\phi^2\right)\left(1+5\phi-20\phi^3+16\phi^5\right)
\ee
The two families are field reflection of one another, and so they are essentially equal for $a$ semi-integer. In Fig.~[1] we plot the $sine$ and $cosine$ potentials for $a=5/2,$ to illustrate their profile. It is interesting to notice that the potentials $V^{1/2}_{\sin}$ and $V^{1/2}_{\cos}$ give rise to the $\phi^3$ theory considered in Ref.~{\cite{vel}} to describe vacuum instability in scalar theories with spontaneous symmetry breaking. These models are of interest to string theory \cite{tac} since they can be used as toy models to describe tachyonic decay of unstable brane \cite{tm}. 

\begin{figure}[ht]
\vspace{1cm}
\includegraphics[{height=08cm,width=6cm,angle=-90}]{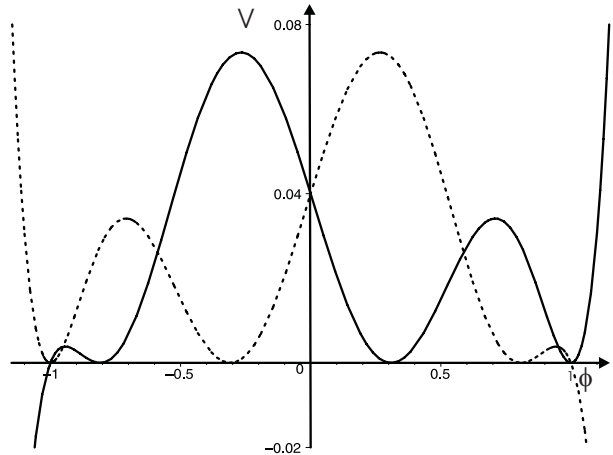}
\vspace{0.3cm}
\caption{The two potentials $V^a_{\sin}$ and $V^a_{\cos},$ plotted for a=5/2 with solid and dashed lines, respectively.}
\end{figure}

The case of $a$ integer is very interesting, and we have decided to show specific investigations for both the $sine$ and $cosine$ families of models below.

\subsection{The $sine$ family of models for $a$ integer}

Here we investigate the family of models given by $V^a_{\sin}$ for the specific case of $a$ being an integer.

\subsubsection{ zeroes of $V_{\sin}$}

The  zeroes of $V^a_{\sin}(\phi)$ imply that $a\arccos\phi=n\pi,\;\; n\in {\mathbb Z}.$ 
We define $k=n+1$, so that $k=1,2,...$. Thus, the  zeroes of ${V}_{\sin}$ are given by
\be
Z_k= \cos \left( \frac{k-1}{a}\pi\right)
\ee

The count of  zeroes runs from $k=1$ to $k=2a+1$, but for $k\leq a+1$ we have the equalities $Z_k=-Z_{a+2-k}$
and $Z_{a+k}=Z_{a+2-k},$ which show that the number of different  zeroes is just $a+1$. The explicit results are: for $a$ even, the  zeroes are
\be
\left\{Z_1\!=\!1, Z_2,\dots, Z_{\frac{a}{2}},Z_{\frac{a}{2}+1}\!=\!0,-Z_{\frac{a}{2}},\dots,-Z_2,-Z_1\!=\!-1 \right\}
\ee
and for $a$ odd, they are
\be
\left\{Z_1=1,\; Z_2,\dots,\;Z_{\frac{a+1}{2}},\;-Z_{\frac{a+1}{2}},\dots,-Z_2,-Z_1=-1 \right\}
\ee
Finally, because ${V}_{\sin}$ is a polynomial for which the  zeroes (and multiplicities) are known, we have:
\begin{itemize}
\item $a$ even:
\be
{V}_{\sin}^a(\phi)=\frac{4^{a/2}}{2a^2}\phi^2 \, \prod_{j=1}^{\frac{a}{2}}
\left(1-\phi^2/Z_j^2\right)^2
\ee
\item $a$ odd:
\be
{V}_{\sin}^a(\phi)=\frac{1}{2a^2}\, \prod_{j=1}^{a}
\left(1-\phi^2/Z_j^2\right)^2
\ee
\end{itemize}
where $Z_j= \cos \frac{j-1}{a}\pi$. It is a very remarkable fact that $V^a_{\sin}(\phi)$ can be written in terms of the Chebyshev
polynomials of second kind:
\bes
\ben
V^a_{\sin}(\phi)&=&\frac1{2a^2}\;U^2_{a-1}(\phi)\;(1-\phi^2)^2
\\
\nonumber
\\
U_a(\phi)&=&\frac{\sin((a+1)\arccos\phi)}{\sin(\arccos\phi)}
\een
\ees
The explicit results of $V^a_{\sin}(\phi)$, for ${a=1,2,3,4,5}$ are given by
\be
V^{1}_{\sin}(\phi)=\frac12(1-\phi^2)^2
\ee
\be
V^{2}_{\sin}(\phi)=\frac12\phi^2\;(1-\phi^2)^2
\ee
\be
V^{3}_{\sin}(\phi)=\frac{8}{9}\left(\frac14-\phi^2\right)^2(1-\phi^2)^2
\ee
\be
V^{4}_{\sin}(\phi)=2\;\phi^2\;\left(\frac12-\phi^2\right)^2(1-\phi^2)^2
\ee
\ben
V^{5}_{\sin}(\phi)&=&\frac{128}{25}\left(\frac{(1+\sqrt{5})^2}{16}-\phi^2\right)^2\nonumber
\\
& &\left(\frac{(1-\sqrt{5})^2}{16}-\phi^2\right)^2\left(1-\phi^2\right)^2
\een
which illustrate this new family of models. We notice that there are two classes of models: for $a$ odd they are $\phi^4-$like potentials -- no zero at the origin -- and for $a$ even they are $\phi^6-$like models -- having a zero at the origin. In Fig.~2 we plot $V^2_{\sin}$ and $V^4_{\sin}$, and in Fig.~3 we plot $V^3_{\sin}$ and $V^5_{\sin}$ to show how they behave as a function of the scalar field $\phi.$ As a curiosity, we remark that the  zeroes of $V^5_{\sin}(\phi)$ occur at $\pm R/2,$ $\pm (1-R)/2,$ and $\pm1,$ where $R=(1+\sqrt{5})/2$ and $1-R$ are respectively the Big and Small Golden Ratios.

\subsubsection{Solutions and  zeroes of ${V}_{\sin}$}

The static solutions of $V_{\sin}(\phi)$ are given by
\be
{\phi}_k(x)= \cos\frac{\theta(x)+(k-1)\pi}{a}
\ee
with $k=1,2,...$ and $\theta(x)$ the principal determination of $\arccos\tanh(x)$, i.e. $\theta\in [0,\pi]$.

Different values of $k$ produce different solutions only if $1\leq
k\leq 2a$. Moreover, if $1\leq k\leq a$, then ${\phi}_{2a+1-k}(x)=\, - \, {\phi}_k(x)$, and this leads to the set of solutions
\be
\left\{{\phi}_1(x), {\phi}_2(x),\dots ,{\phi}_a(x), -{\phi}_a(x),\dots, -{\phi}_2(x),-{\phi}_1(x) \right\}
\ee

We notice that for $-\infty \leq x\leq \infty$ we have $-1\leq \tanh x\leq 1$ and so $\pi\geq\theta\geq0$. Thus, $x\to -\infty$ implies
$\theta \to \pi$ and $x\to \infty$ implies $\theta\to 0$. We use these results to see that from $k=1$ to $k=a$, the solution ${\phi}_k(x)$
interpolates between the zero $Z_{k+1}$ and the zero $Z_{k}$, all the double  zeroes are reached, and these are {\it topological defects.} The solutions ${\phi}_{a+k}(x)$ travel in the opposite direction, and they are the associated anti-defects.

\begin{figure}[ht]
\vspace{1cm}
\includegraphics[{height=08cm,width=6cm,angle=-90}]{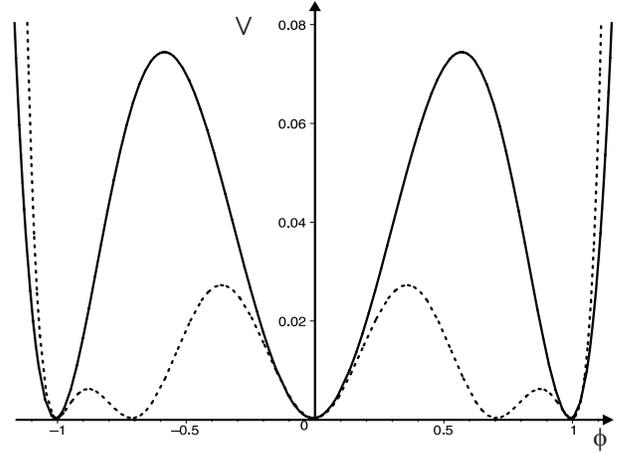}
\vspace{0.3cm}
\caption{Plots of $V^2_{\sin}(\phi)$ and $V^4_{\sin}(\phi),$ depicted with solid and dashed lines, respectively. These potentials belong
to the $\phi^6$-like family of models.}
\end{figure}

\begin{figure}[ht]
\vspace{1cm}
\includegraphics[{height=08cm,width=6cm,angle=-90}]{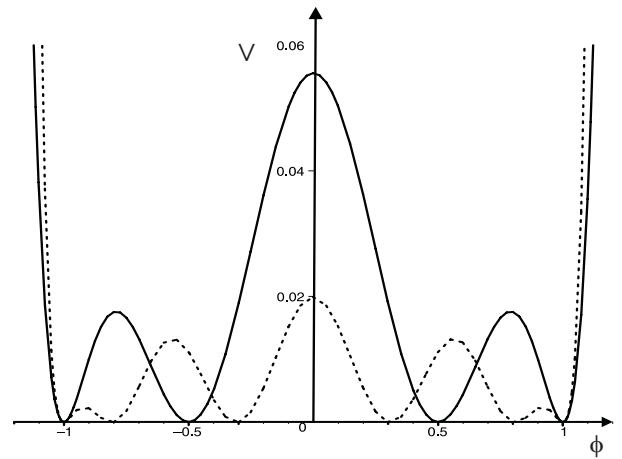}
\vspace{0.3cm}
\caption{Plots of $V^3_{\sin}(\phi)$ and $V^5_{\sin}(\phi),$ depicted with solid and dashed lines, respectively. These potentials belong to the $\phi^4$-like family of models.}
\end{figure}

\subsection{The $cosine$ family of models for $a$ integer}

Here we investigate the family of models given by $V^a_{\cos}$ for the specific case of $a$ being  an integer.

\subsubsection{The  zeroes of ${V}_{\cos}$}

The  zeroes of $V^a_{\cos}(\phi)$ imply that $a\arccos\phi=\frac{\pi}{2}+n\pi, n\in {\mathbb Z}$. Like before, we define $k=n+1$, so that $k=1,2,...$. Thus, apart from $\pm 1$, the  zeroes of $V_{\cos}$ are
\be
Z_k= \cos \left( \frac{2k-1}{2a}\pi\right)
\ee

The count of  zeroes runs from $k=1$ to $k=2a$, but for $k\leq a$ the equalities $Z_k=-Z_{a-k+1}$ and $Z_{a+k}=Z_{a-k+1}$ show that the number of different  zeroes is just $a$ (plus the two simple  zeroes $\pm 1$). For $a$ even, the  zeroes are
\be
\left\{ 1, Z_1, Z_2,\dots, Z_{\frac{a}{2}}, -Z_{\frac{a}{2}}, \dots,
-Z_2,-Z_1, -1 \right\}
\ee
and for $a$ odd, they are
\be
\left\{1,Z_1,Z_2,\dots,Z_{\frac{a-1}{2}},Z_{\frac{a+1}{2}}\!\!=\!0,\!-Z_{\frac{a-1}{2}},\dots,\!-Z_2,\!-Z_1,\!-1\right\}
\ee

Finally, because ${V}_{\cos}$ is a polynomial for which the  zeroes (and multiplicities) are known, we have:
\begin{itemize}
\item $a$ even:
\be
{V}_{\cos}^a(\phi)=\frac{4^{a-1}}{2a^2}(1-\phi^2) \, \prod_{j=1}^{\frac{a}{2}}
\left( \phi^2-Z_j^2\right)^2
\ee
\item $a$ odd:
\be
{V}_{\cos}^a(\phi)=\frac{4^{a-1}}{2a^2}\phi^2 (1-\phi^2) \,
\prod_{j=1}^{\frac{a-1}{2}} \left( \phi^2-Z_j^2\right)^2
\ee
\end{itemize}
where $Z_j= \cos \frac{2j-1}{2a}\pi$. As for the $sine$ potentials, the $V^a_{\cos}(\phi)$ potentials are also given by Chebyshev polynomials,
in this case of the first kind:
\bes\ben
V^a_{\cos}(\phi)=\frac1{2a^2}\;T^2_a(\phi)\;(1-\phi^2)
\\
\nonumber
\\
T_a(\phi)=\cos(a\arccos\phi)
\een\ees
Note the relationship between $T_a(\phi)$ and the deformation function introduced in Eq.~(\ref{df}).
 
The explicit forms of potentials $V^a_{\cos}(\phi)$ for ${a=1,2,3,4,5}$ are given by
\be
V^{1}_{\cos}(\phi)=\frac12\phi^2\;(1-\phi^2)
\ee
\be
V^{2}_{\cos}(\phi)=\frac12\left(\frac12-\phi^2\right)^2(1-\phi^2)
\ee
\be
V^{3}_{\cos}(\phi)=\frac89\phi^2\;\left(\frac34-\phi^2\right)^2(1-\phi^2)
\ee
\be
V^{4}_{\cos}(\phi)=2\left(\frac{2+\sqrt{2}}{4}-\phi^2\right)^2\left(\frac{2-\sqrt{2}}{4}-\phi^2\right)^2(1-\phi^2)
\ee
\ben
V^{5}_{\cos}(\phi)&=&\frac{128}{25}\;\phi^2\;\left(\frac{10+2\sqrt{5}}{16}-\phi^2\right)^2\nonumber
\\
& &\left(\frac{10-2\sqrt{5}}{16}-\phi^2\right)^2(1-\phi^2)
\een
which illustrate this new family of models. We notice that there are two classes of models: for $a$ odd they are inverted $\phi^4-$like models, and for $a$ even they are inverted $\phi^6-$like models. In Fig.~4 we plot $V^2_{\cos}$ and $V^4_{\cos},$ and in Fig.~5 we plot $V^3_{\cos}$ and $V_{\cos}^5$ to show how they behave as a function of the scalar field $\phi.$

\begin{figure}[ht]
\vspace{1cm}
\includegraphics[{height=08cm,width=6cm,angle=-90}]{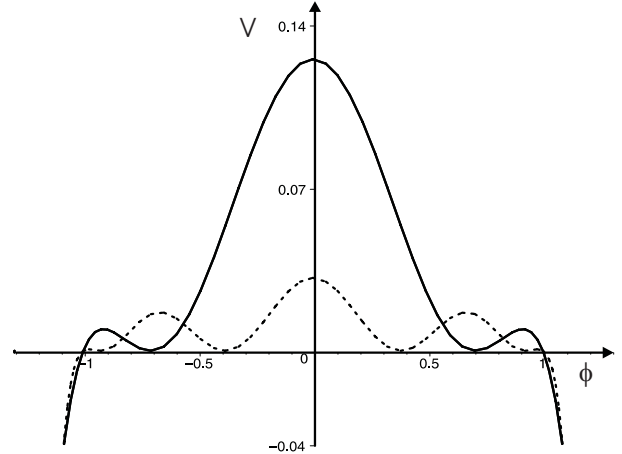}
\vspace{0.3cm}
\caption{Plots of $V^2_{\cos}(\phi)$ and $V^4_{\cos}(\phi),$ depicted with solid and dashed lines, respectively. These potentials
belong to the inverted $\phi^6$-like family of models.}
\end{figure}

\begin{figure}[ht]
\vspace{1cm}
\includegraphics[{height=08cm,width=6cm,angle=-90}]{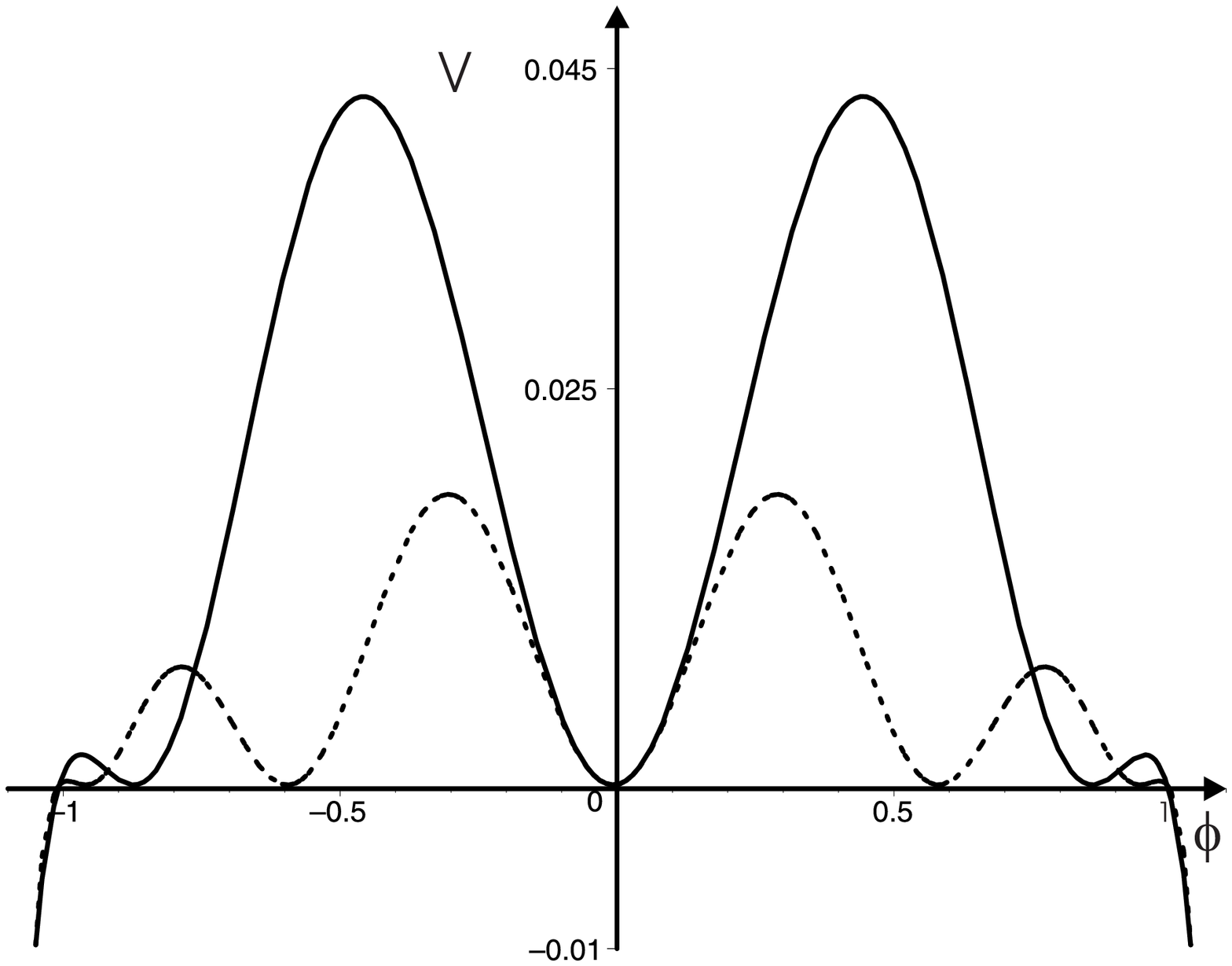}
\vspace{0.3cm}
\caption{Plots of $V^3_{\cos}(\phi)$ and $V^5_{\cos}(\phi),$ depicted with solid and dashed lines, respectively. These potentials belong
to the inverted $\phi^4$-like family of models.}
\end{figure}

\subsubsection{Solutions and  zeroes of ${V}_{\cos}$}

The static solutions of $V_{\cos}(\phi)$ are given by
\be
{\phi}_k(x)= \cos\frac{\theta(x)+\frac{2k-1}{2}\pi}{a}
\ee
with $k=1,2,...,$ and $\theta(x)$ the principal determination of $\arccos \tanh(x-x_0)$, i.e. $\theta\in [0,\pi]$.

Different values of $k$ produce different solutions only if $1\leq k\leq 2a$. Moreover, if $1\leq k< a$, then
${\phi}_{2a-k}(x)= - {\phi}_k(x)$ and we get the set of solutions ${\phi}_1(x),$ ${\phi}_2(x),$ $\dots,$ ${\phi}_{a-1}(x),$
${\phi}_a(x),$ $-{\phi}_{a-1}(x),$ $\dots,$ $-{\phi}_2(x),$ $-{\phi}_1(x),$ ${\phi}_{2a}(x).$

We see that for $-\infty\leq x\leq\infty$ we have $-1\leq\tanh x\leq1$ and so $\pi\geq\theta\geq0$.
Thus, $x\to-\infty$ gives $\theta\to\pi$ and $x\to\infty$ gives $\theta\to0$. From $k=1$ to $k=a-1$, the solution ${\phi}_k(x)$
interpolates between the zero $Z_{k+1}$ and the zero $Z_{k}$, all the double  zeroes are reached, and these are {\it topological
defects}. The solutions ${\phi}_{a+k}(x)$ travel in the opposite direction, and they are the associated anti-defects. Moreover,
${\phi}_a$ and ${\phi}_{2a}$ are {\it non-topological defects}. ${\phi}_a(x)$ starts at $x=-\infty$ from the zero
$Z_{a+1}=-Z_1$, at $x=0$ (more generally at $x=x_0$) reaches ${\phi}_a(x_0)=-1$, and finally at $x=\infty$ the solution
arrives to $Z_a$. But $Z_a=-Z_1$ again. Identical behavior appears for ${\phi}_{2a}(x)$ on the  zeroes $Z_{2a}=Z_1$, and $Z_{2a+1}\equiv Z_1$.

\section{Superpotentials}

It is sometimes possible to introduce superpotentials. They are functions of the scalar field, $W=W(\phi),$ and allow writing
\be
V(\phi)=\frac12\left(\frac{dW}{d\phi}\right)^2
\ee

Since for the $sine$ family of potentials, $V^a_{\sin}(\phi),$ all members with $a$ integer are non-negative, we can introduce the corresponding superpotentials. They are obtained with
\be
{V}_{\sin}(\phi)=\frac12\left(\frac{dW_{\sin}(\phi)}{d\phi}\right)^2
\ee
They have the form, for $a^2\neq4$
\ben
W_{\sin}(\phi)\!=\!&\pm&\frac{1}{a^2(a^2-4)}[( a^2(1-\phi^2)-2)\cos(a\,\arccos\phi)\nonumber\\
&-& 2a \, \phi\sqrt{1-\phi^2} \, \sin( a \,\arccos\phi)]
\een
or
\ben
W_{\sin}(\phi)\!=\!&\pm&\frac1{a^2(a^2-4)}[(a^2(1-\phi^2)-2)T_a(\phi)\nonumber\\
&-&2a\phi(1-\phi^2)U_{a-1}(\phi)]
\een
and for $a^2= 4$ we have
\be
W_{\sin}(\phi)=\pm \frac{1}{4} \, \phi^2 \left( 2-\phi^2\right)
\ee

The presence of superpotentials in general eases calculation; in particular, this is the case for the energy associated with the corresponding static solutions. For the kink solutions of the $sine$ family of models, the energies can be calculated with
\be
E_{\sin}({\phi_k})=\left| W_{\sin}(Z_{k})-
W_{\sin}(Z_{k+1})\right|
\ee
for $1\leq k\leq 2a.$ The presence of superpotentials shows that these models are in fact bosonic portions of more general, globally
supersymmetric theories, which can be written in the usual way \cite{susy,susy1}. We recall that global supersymmetries can be made local,
as it happens in supergravity theories, which can also support defect structures -- see, for instance, Ref.~{\cite{CS}}.

\section{Stability and zero modes}

To investigate linear stability of the scalar field solutions Eq.~(\ref{sol1}), we consider
\be
{\phi}(x,t)={\phi}(x)+\sum_{n}\eta(x)\cos(w_n t)
\ee
We substitute this into the time-dependent equation of motion
\be
\frac{\partial^2\phi}{\partial t^2}-\frac{\partial^2\phi}{\partial x^2}+\frac{dV}{d\phi}=0
\ee
and consider the case of small fluctuations $\eta_n (x)$ about the classical field ${\phi}(x)$ to obtain
\be\label{sch}
\left(-\frac{d^2}{dx^2}+U(x)\right)\eta_n (x)=w_n^2\;\eta_n (x)
\ee
where the potential of this Sch\"odinger-like equation is
\be\label{u1}
U(x)=\left.\frac{d^2
{V}}{d\phi^2}\right|_{\phi={\phi}(x)}
\ee
The equation (\ref{sch}) has at least one bound state, the bosonic zero mode that is present due to translational invariance. It is given by
\be
\eta_0 (x)= \frac{d{\phi}(x)}{dx}
\ee
From Eq.~(\ref{sol1}) we obtain
\be
\eta_0(x)=\frac{1}{a}\;
{\rm sech}(x)\, \sin \left(\frac{\theta(x)+m\pi}{a}\right)
\ee
where $\theta(x)$ is the principal determination of $\arccos(\tanh(x))$. For the above solutions, zero modes have nodes only for semi-integer $m=\frac{2k-1}{2}$, with $k=a$ and $k=2a$. Yet, from Eq.~(\ref{u1}) for the solutions (\ref{sol1}), after a highly non trivial calculation, we have the potentials
\ben
U({\phi}_k(x))&=&1-\left(2+\frac{1}{a^2}\right){\rm sech}^2(x) +\nonumber\\
&&\frac3a {\rm cot}\left(\frac{\theta(x)+N(k)\pi}{a}\right)\tanh(x)\;{\rm sech}(x){\;\;\;\;\;\;\;}
\een
where $N(k)=k-1$, for the solutions ${\phi}_k(x)$ of ${V}_{sin}$, and $N(k)=(2k-1)/2$ for the solutions of ${V}_{cos}.$

The explicit forms of the potentials $V(\phi)$ and quantum mechanical potentials $U(x)$ for $a=1,2$ are shown below. For $a=1$ we have
\be
{V}_{\sin}^1=\frac12(1-\phi^2)^2
\ee
\be
{U}_{\sin}^1({\phi}_1)={U}_{\sin}^1({\phi}_2)=4-6\;{\rm sech}^2(x)
\ee
and
\be
{V}_{\cos}^1=\frac12\phi^2(1-\phi^2)
\ee
\be
{U}_{\cos}^1({\phi}_1)={U}_{\cos}^1({\phi}_2)=1-6\;{\rm sech}^2(x)
\ee
For $a=2$ we have
\be
{V}_{\sin}^2=\frac{1}{2}\phi^2 \left( \frac{1}{2}-\phi^2\right)^2 
\ee
\be
{U}_{\sin}^2({\phi}_1)={U}_{\sin}^2({\phi}_3)=\frac{1}{4}\left(10+6\;{\rm sech}(x)-15\,{\rm sech}^2(x)\right)
\ee
\be
{U}_{\sin}^2({\phi}_2)={U}_{\sin}^2({\phi}_4)=\frac{1}{4}\left(10-6\;{\rm sech}(x)-15\,{\rm sech}^2(x)\right)
\ee
and 
\be
{V}_{\cos}^2=\frac{1}{2}(1-\phi^2)\left(\frac{1}{2}-\phi^2\right)^2
\ee
\be
{U}_{\cos}^2({\phi}_1)={U}_{\cos}^2({\phi}_3)=\frac{1}{4}\left(4+6\;{\rm sech}(x)-15\;{\rm sech}^2(x)\right)
\ee
\be
{U}_{\cos}^2({\phi}_2)={U}_{\cos}^2({\phi}_4)=\frac{1}{4}\left(4-6\;{\rm sech}(x)-15\;{\rm sech}^2(x)\right)
\ee

We notice that the zero modes for topological defects are zero energy eigenfunctions of the Hessian for both ${V}_{\sin}$ and 
${V}_{\cos}$ and have no nodes. The Hessian is a one-dimensional second-order Schr\"odinger operator, and by a general theorem, the
ground state of such operator has no nodes. Therefore, all the other eigenfunctions have positive energy and topological defects are stable.
The zero modes for non-topological defects are also zero energy eigenfunctions of the Hessian for ${V}_{\cos}$, but they
have a node. Therefore, the ground state in this case is a negative energy eigenfunction, indicating that non-topological defects are unstable.

The above study of stability shows the presence of new quantum-mechanical potentials, which is of direct interest to supersymmetric quantum mechanics and can be explored standardly, with the usual procedure of factorization \cite{susyqm}.

\section{Final comments}

In this work we have built two distinct families of scalar field models, described by the potentials $V^a_{\sin}(\phi)$ and $V^a_{\cos}(\phi),$
given by the general expressions in (\ref{vs1}) and (\ref{vc1}). For these models, we have found all the static solutions, and we have studied their stabilities carefully. As we have shown, the $V^a_{\sin}(\phi)$ models are generalizations of the $\phi^4$ and $\phi^6$ models, for $a$ odd and even, respectively, and the $V^a_{\cos}(\phi)$ generalizes the inverted $\phi^4$ and inverted $\phi^6$ models, for $a$ odd and even, respectively. These models are new, of increasing power in the scalar field, and they are all solved with explicit static solutions obtained very directly from the procedure for deformed defect set forward in Ref.~{\cite{dd}}.   

The present investigation offers a systematic way to find defect solutions in models described by real scalar field with general polynomial interactions. The simplicity of the procedure poses several issues of interest to high energy physics and condensed matter. In high energy physics, in particular, we are now focusing on applications to tachyons and branes, and on extensions to models described by two real scalar fields \cite{two}, or by a single complex field \cite{bms}. In connection with the semiclassical program of Ref.~{\cite{2dqft}}, we are now examining the possibility of extending the present work to models defined in finite volume.

The authors would like to thank the Brazilian and Spanish agencies and projects, respectively CAPES, CNPq, PADCT/CNPq and PRONEX/CNPq/FAPESQ, and DGICYT and JCyL for partial financial support. JMG is grateful to the Departamento de F\'\i sica, Universidade Federal da Para\'\i ba, Jo\~ao Pessoa,
for warm hospitality.


\end{document}